# Determinism and Indeterminism

Robert C. Bishop
Faculty of Philosophy
University of Oxford

Determinism is a rich and varied concept. At an abstract level of analysis, Jordan Howard Sobel (1998) identifies at least ninety varieties of what determinism could be like. When it comes to thinking about what deterministic laws and theories in physical sciences might be like, the situation is much clearer. There is a criterion by which to judge whether a law–expressed as some form of equation–is deterministic. A theory would then be deterministic just in case all its laws taken as a whole were deterministic. In contrast, if a law fails this criterion, then it is indeterministic and any theory whose laws taken as a whole fail this criterion must also be indeterministic. Although it is widely believed that classical physics is deterministic and quantum mechanics is indeterministic, application of this criterion yields some surprises for these standard judgments.

**Framework for Physical Theories**
Laws and theories in physics are formulated in terms of dynamical or evolution equations. These equations are taken to describe the change in time of the relevant variables characterizing the system in question. Additionally, a complete specification of the initial state referred to as the initial conditions for the system and/or a characterization of the boundaries for the system known as the boundary conditions must also be given. A state is taken to be a description of the values of the variables characterizing the system at some time $t$. As a simple example of a classical model, consider a cannon firing a ball. The initial conditions would be the initial position and velocity of the ball as it left the mouth of the cannon. The evolution equation plus these initial conditions would then describe the path of the ball.

Much of the analysis of physical systems takes place in what is called state space, an abstract mathematical space composed of the variables required to fully specify the state of a system. Each point in this space then represents a possible state of the system at a particular time $t$ through the values these variables take on at $t$. For example, in many typical dynamical models–constructed to satisfy the laws of a given theory–the position and momentum serve as the coordinates, so the model can be studied in state space by following its trajectory from the initial state ($q_o$, $p_o$) to some final state ($q_f$, $p_f$). The evolution equations govern the path–the history of state transitions–of the system in state space.

However, note that there are important assumptions being made here. Namely, that a state of a system is characterized by the values of the crucial variables and that a physical state corresponds to a point in state space through these values. This cluster of assumptions can be called the faithful model assumption. This assumption allows one to develop mathematical models for the evolution of these points in state space and such models are taken to represent (perhaps through a complicated relation) the physical systems of interest. In other words, one assumes that one's mathematical models are faithful representations of physical systems and that the state space is a faithful representation of the space of physically genuine possibilities for the system in question. Hence, one has the connection between physical systems and their laws and models, provided the latter are faithful. It then remains to determine whether these laws and models are deterministic or not.

**Laplacean Determinism**
Clocks, cannon balls fired from cannons and the solar system are taken to be paradigm examples of deterministic systems in classical physics. In the practice of physics, one is able to give a very general and precise description of deterministic systems. For definiteness the focus here is on classical particle mechanics, the inspiration for Pierre Simon Laplace's famous description:

> We ought to regard the present state of the universe as the effect of its antecedent state and as the cause of the state that is to follow. An intelligence knowing all the forces acting in nature at a given instant, as well as the momentary positions of all things in the universe, would be able to comprehend in one single formula the motions of the largest bodies as well as the lightest atoms in the world...to it nothing would be uncertain, the future as well as the past would be present to its eyes.(translation from Ernst Nagel 1961, pp. 281-282)

Given all the forces acting on the particles composing the universe along with their exact positions and momenta, then the future behavior of these particles is, in principle, completely determined.
   Two historical remarks are in order here. First, Laplace's primary aim in this famous passage was to contrast the concepts of probability and certainty. Second, Gottfried Wilhelm Leibniz (1924, p. 129) articulated this same notion of inevitability in terms of particle dynamics long before Laplace. Nevertheless, it was the vision that Laplace articulated that has become a paradigm example for determinism in physical theories.
   This vision may be articulated in the modern framework as follows. Suppose that the physical state of a system is characterized by the values of the positions and momenta of all the particles composing the system at some time $t$. Furthermore, suppose that a physical state corresponds to a point in state space (invoking the faithful model assumption). One can then develop deterministic mathematical models for the evolution of these points in state space. Some have thought that the key feature characterizing this determinism was that given a specification of the initial state of a system and the evolution equations governing its states, in principle it should be possible to predict the behavior of the system for any time (recall Laplace's contrast between certainty and probability). Although prima facie plausible, such a condition is neither necessary nor sufficient for a deterministic law because the relationship of predictability to determinism is far too weak and subtle.
   Rather, the core feature of determinism is the following condition:

> <u>Unique Evolution</u>: A given state is always followed (and preceded) by the same history of state transitions.

This condition expresses the Laplacean belief that systems described by classical particle mechanics will repeat their behaviors exactly if the same initial and boundary conditions are specified. For example the equations of motion for a frictionless pendulum will produce the same solution for the motion as long as the same initial velocity and initial position are chosen. Roughly speaking, the idea is that every time one returns the mathematical model to the same initial state (or any state in the history of state transitions), it will undergo the same history of transitions from state to state and likewise for the target system. In other words the evolution will be unique given a specification of initial and boundary conditions. Note that as formulated, unique evolution expresses state transitions in both directions (future and past). It can easily be recast to allow for unidirectional state transitions (future only or past only) if desired.

**Unique Evolution**
Unique evolution is the core of the Laplacean vision for determinism (it lies at the core of Leibniz's statement as well). Although a strong requirement, it is important if determinism is to be meaningfully applied to laws and theories. Imagine a typical physical system $s$ as a film. Satisfying unique evolution means that if the film is started over and over at the same frame (returning the system to the same initial state), then $s$ will repeat every detail of its total history over and over again and identical copies of the film would produce the same sequence of pictures. So if one always starts Jurassic Park at the beginning frame, it plays the same. The tyrannosaurus as antihero always saves the day. No new frames are added to the movie. Furthermore, if one were to start with a different frame, say a frame at the middle of the movie, there is still a unique sequence of frames.

By way of contrast, suppose that returning $s$ to the same initial state produced a different sequence of state transitions on some of the runs. Consider a system $s$ to be like a device that spontaneously generates a different sequence of pictures on some occasions when starting from the same initial picture. Imagine further that such a system has the property that simply by choosing to start with any picture normally appearing in the sequence, sometimes the chosen picture is not followed by the usual sequence of pictures. Or imagine that some pictures often do not appear in the sequence, or that new ones are added from time to time. Such a system would fail to satisfy unique evolution and would not qualify as deterministic.

More formally, one can define unique evolution in the following way. Let $S$ stand for the collection of all systems sharing the same set $L$ of physical laws and suppose that $P$ is the set of relevant physical properties for specifying the time evolution of a system described by $L$:

> A system $s \in S$ exhibits unique evolution if and only if every system $s' \in S$ isomorphic to $s$ with respect to $P$ undergoes the same evolution as $s$.

**Two Construals of Unique Evolution**
Abstracting from the context of physical theories for the moment, unique evolution can be given two construals. The first construal is as a statement of causal determinism, that every event is causally determined by an event taking place at some antecedent time or times. This reading of unique evolution fits nicely with how a number of philosophers conceive of metaphysical, physical and psychological determinism as theses about the determination of events in causal chains, where there is a flow from cause to effect, if you will, that may be continuous or have gaps. The second construal of unique evolution is as a statement of difference determinism characterized by William James as "[t]he whole is in each and every part, and welds it with the rest into an absolute unity, an iron block, in which there can be no equivocation or shadow of turning"(1956, p. 150). This reading of unique evolution maintains that a difference at any time requires a difference at every time.

These two construals of unique evolution are different. For example, consider a fast-starting series of causally linked states (Sobel 1998, p. 89), where every state in the series has an earlier determining cause, but the series itself has no antecedent deterministic cause (its beginning–the first state–is undetermined by prior events or may have a probabilistic cause) and no state in the series occurs before a specified time. The principle that every event has an earlier cause would fail for a fast-starting series as a whole though it would hold for the events within such a series. This would be an example where causal determinism failed, but where difference determinism would still hold.

However, the causal construal of unique evolution is unsatisfactory. Concepts like *event* or *causation* are vague and controversial. One might suggest explicating causal determinism in terms of the laws *L* and properties *P*, but concepts like *event* and *cause* are not used in most physical theories (at least not univocally). In contrast, unique evolution fits the idea of difference determinism: any difference between *s* and *s*ℕ is reflected by different histories of state transitions. This latter construal of unique evolution only requires the normal machinery of the theoretical framework sketched earlier in order to cash out these differences and so avoids controversies associated with causal determinism.

**Determinism in Classical Mechanics**
Most philosophers take classical mechanics to be the archetype of a deterministic theory. Prima facie Newton's laws satisfy unique evolution (see Newtonian Mechanics). After all, these are ordinary differential equations and one has uniqueness and existence proofs for them. Furthermore, there is at least some empirical evidence that macroscopic objects behave approximately as these laws describe. Still, there are some surprises and controversy regarding the judgment that classical mechanics is a deterministic theory.

For example, as Keith Hutchinson (1993, p. 320) notes, if the force function varies as the square root of the velocity, then a specification of the initial position and velocity of a particle does not fix a unique evolution of the particle in state space (indeed, the particle can sit stationary for an arbitrary length of time and then spontaneously begin to move). Hence, such a force law is not deterministic. There are a number of such force functions consistent with Newton's laws, but that fail to satisfy unique evolution. Therefore, the judgment that classical mechanics is a deterministic theory is false.

<u>Newtonian Gravity</u>. One might think that the set of force functions leading to violations of unique evolution represents an unrealistic set so that all force laws of classical mechanics really are deterministic. However, worries for determinism await one even in the case of point particles interacting under Isaac Newton's force of gravity, the paradigm case of determinism that Laplace had in mind.

In 1897 the French mathematician Paul Painlevé conjectured that a system of point-particles interacting only under Newton's force of gravity could all accelerate to spatial infinity within a finite time interval. (The source of the energy needed for this acceleration is the infinite potential well associated with the inverse-square law of gravitation.) If particles could disappear to 'spatial infinity,' then unique evolution would break down because solutions to the equations of motion no longer would be guaranteed to exist. Painlevé's conjecture was proven by Zhihong Xia (1992) for a system of five point-masses.

Though provocative, these results are not without controversy. For example, there are two interesting possibilities for interpreting the status of these particles that have flown off to spatial infinity. On the one hand, one could say the particles have left the universe and now has some indefinite properties. On the other hand, one could say that the particles no longer exists. Newton's mechanics is silent on this interpretive question. Furthermore, are events such as leaving the universe to be taken as predictions of Newton's gravitational theory of point particles, or as indications that the theory is breaking down because particle position becomes undefined? Perhaps such behavior is an artifact of a spatially infinite universe. If the universe is finite, particle positions are always bounded and such violations of unique evolution are not possible.

<u>Diagnosis</u>. Other failures of unique evolution in classical mechanics can be found in John Earman's (1986) survey. What is one to say, then, about the uniqueness and existence theorems for the equations of motion, the theorems that appear so suggestive of unique evolution? The root problem of these failures to satisfy unique evolution can be traced back to the fact that one's mathematical theorems only guarantee existence and uniqueness locally in time. This means that the equations of motion only have unique solutions for some interval of time. This interval might be short and, as time goes on, the interval of time for which such solutions exist might get shorter or even shrink to zero in such a way that after some period solutions cease to exist. So determinism might hold locally, but this does not guarantee determinism must hold globally.

**Determinism in Special and General Relativity**
Special relativity provides a much more hospitable environment for determinism. This is primarily due to two features of the theory: (1) no process or signal can travel faster than the speed of light, and (2) the spacetime structure is static. The first feature rules out unbounded-velocity systems, while the second guarantees there are no singularities in spacetime. Given these two features, global existence and uniqueness theorems can be proven for cases like source-free electromagnetic fields so that unique evolution is not violated when appropriate initial data are specified on a space-like hypersurface. Unfortunately, when electromagnetic sources or gravitationally interacting particles are added to the picture, the status of unique evolution becomes much less clear.

In contrast, general relativity presents problems for guaranteeing unique evolution. For example, there are spacetimes for which there are no appropriate specifications of initial data on space-like hypersurfaces yielding global existence and uniqueness theorems. In such spacetimes, unique evolution is easily violated. Furthermore, problems for unique evolution arise from the possibility of naked singularities (singularities not hidden behind an event horizon). One way a singularity might form is from gravitational collapse. The usual model for such a process involves the formation of an event horizon (i.e., a black hole). Although a black hole has a singularity inside the event horizon, outside the horizon at least determinism is okay, provided the spacetime supports appropriate specifications of initial data compatible with unique evolution. In contrast, a naked singularity has no event horizon. The problem here is that anything at all could pop out of a naked singularity, violating unique evolution. To date, no general, convincing forms of hypotheses ruling out such singularities have been proven (so-called cosmic censorship hypotheses).

**Determinism in Quantum Mechanics**
In contrast to classical mechanics philosophers often take quantum mechanics to be an indeterministic theory. Nevertheless, so-called pilot-wave theories pioneered by Louis de Broglie and David Bohm are explicitly deterministic while still agreeing with experiments. Roughly speaking, this family of theories treats a quantum system as consisting of both a wave and a particle. The wave evolves deterministically over time according to the Schrödinger equation and determines the motion of the particle. Hence, the particle's motion satisfies unique evolution. This is a perfectly coherent view of quantum mechanics and contrasts strongly with the more orthodox interpretation. The latter takes the wave to evolve deterministically according to Schrödinger's equation and treats particle-like phenomena indeterministically in a measurement process (such processes typically violate unique evolution because the particle system can be in the same state before measurement, but still yield many different outcomes after measurement). Pilot-wave theories show that quantum mechanics need not be indeterministic.

**Deterministic Chaos**
Some philosophers have thought that the phenomenon of deterministic chaos–the extreme sensitivity of a variety of classical mechanics systems such that roughly even the smallest change in initial conditions can lead to vastly different evolutions in state space–might actually show that classical mechanics is not deterministic. However, there is no real challenge to unique evolution here as each history of state transitions in state space is still unique to each slightly different initial condition.

Of course, classical chaotic systems are typically considered as if there is no such thing as quantum mechanics. But suppose one considers a combined system such that quantum mechanics is the source of the small changes in initial conditions for one's classical chaotic system? Would such a system fail to satisfy unique evolution? The worry here is that since there is no known lower limit to the sensitivity of classical chaotic systems, nothing can prevent the possibility of such systems amplifying a slight change in initial conditions due to a quantum event so that the evolution of the classical chaotic system is dramatically different than if the quantum event had not taken place. Indeed, some philosophers argue that unique evolution must fail in such circumstances.

However, such sensitivity arguments depend crucially on how quantum mechanics itself and measurements are interpreted as well as on where the cut is made distinguishing between what is observed and what is doing the observing (e.g., is the classical chaotic system serving as the measuring device for the quantum change in initial conditions?). Although considered abstractly, sensitivity arguments do correctly lead to the conclusion that quantum effects can be amplified by classical chaotic systems; they do not automatically render one's classical plus quantum system indeterministic. Furthermore, applying such arguments to concrete physical systems shows that the amplification process may be severely constrained. For example investigating the role of quantum effects in the process of chaos in the friction of sliding surfaces indicates that quantum effects might be amplified by chaos to produce a difference in macroscopic behavior only if the fluctuations are large enough to break molecular bonds and are amplified quickly enough.

**Broader Implications**
Finally, what of broader implications of determinism and indeterminism in physical theories? Debates about free will and determinism are one place where the considerations in this entry might be relevant. One of the most discussed topics in this regard is the consequence argument, which may be put informally as follows: If determinism is true, then a person's acts are consequences of laws and events in the remote past. But what went on before a person was born is not up to the person and neither are the laws. Therefore, the consequences of these laws and events–including a person's present acts–are not up to the person. Whether or not the relevant laws satisfy unique evolution is one factor in the evaluation of this argument.

What of broader philosophical thinking about psychological determinism or the thesis that the universe is deterministic? For the former, it looks difficult to make any connection at all. One simply does not have any theories in the behavioral sciences that are amenable to analysis under the criterion of unique evolution. Indeed, attempts to apply the criterion in psychology do not lead to clarification of the crucial issues (Bishop 2002).

With regards to the universe, it has been common practice since the seventeenth century for philosophers to look to their best scientific theories as guides to the truth of determinism. As we have seen, our current best theories in physics are remarkably unclear about the truth of determinism in the physical sciences, so the current guides do not appear to be so helpful. Even if the best theories were clear on the matter of determinism in their province, there is a further problem awaiting their application to metaphysical questions about the universe as a whole. Recall the crucial faithful model assumption. In many contexts this assumption is fairly unproblematic. However, if the system in question is nonlinear–that is to say, has the property that a small change in the state or conditions of the system is not guaranteed to result in a small change in the system's behavior–this assumption faces serious difficulties (indeed, a strongly idealized version of the assumption, the perfect model scenario is needed but also runs into difficulties regarding drawing conclusions about the systems one is modeling). Since the universe is populated with such systems–indeed, it is likely to be nonlinear itself–one's purchase on applying our best laws and theories to such systems or the universe as a whole to answer the large metaphysical question about determinism is quite problematic.

**Bibliography**
Relevant historical material on determinism:
James, William. "The Dilemma of Determinism." In <u>The Will to Believe and Other Essays in Popular Philosophy and Human Immortality</u>. New York: Dover Publications, 1956.
Laplace, Pierre Simon de <u>A Philosophical Essay on Probabilities</u>. Translated by Frederick Wilson Truscott and Frederick Lincoln Emory. New York: Dover Publications, 1814/1951.
Leibniz, Gottfried, Wilhelm "Von dem Verhängnisse." In <u>Hauptschriften zur Grundlegung der Philosophie, Vol. 2</u>. Edited by Ernst Cassirer and Artur Buchenau. Leipzig: Meiner, 1924.
Nagel, Ernst. <u>The Structure of Science: Problems in the Logic of Scientific Explanation</u>. New York: Harcourt, Brace, and World, 1961.
Sobel, Jordan. Howard <u>Puzzles for the Will: Fatalism, Newcomb and Samarra, Determinism and Omniscience</u>. Toronto: University of Toronto Press, 1998.

Laplace's vision expressed in the modern framework of physical theories, as well as discussions of chaos, prediction and determinism, may be found in:
Bishop, Robert C. "On Separating Predictability and Determinism." <u>Erkenntnis</u> 58(2003):169-188.
Bishop, Robert. C. and Kronz, Frederick M. "Is Chaos Indeterministic?" In <u>Language, Quantum, Music: Selected Contributed Papers of the Tenth International Congress of Logic, Methodology & Philosophy of Science, Florence, August 1995</u>. Edited by Maria Luisa Dalla Chiara, Roberto Giuntini and Federico Laudisa. Boston:

    Kluwer Academic Publishers, 1999.

Hobbs, Jesse. "Chaos and Indeterminism," <u>Canadian Journal of Philosophy</u> 21(1991):141-164.

Stone, M. A. "Chaos, Prediction and Laplacean Determinism." <u>American Philosophical Quarterly</u> 26(1989):123-131.

There are a number of able discussions of problems for determinism in physical theories. The following all discuss classical physics; see Earman (1986, 2004) for discussions of determinism in relativistic physics:

Earman, John. <u>A Primer on Determinism</u>. Dordrecht, The Netherlands: D. Reidel Publishing, 1986.

Earman, John. " Determinism: What We Have Learned and What We Still Don't Know." In <u>Freedom and Determinism</u>. Edited by Joseph Keim Campbell, Michael O'Rourke and David Shier. Cambridge, Mass: MIT Press, 2004, pp. 21-46.

Hutchinson, Keith. "Is Classical Mechanics Really Time-Reversible and Deterministic?" <u>British Journal for the Philosophy of Science</u> 44(1993):307-323.

Xia, Zhihong "The Existence of Noncollision Singularities in Newtonian Systems." <u>Annals of Mathematics</u> 135 (3) (1992):411-468.

Uniqueness and existence proofs for differential equations are discussed by:

Arnold, V. I. <u>Geometrical Methods In The Theory Of Ordinary Differential Equations, 2$^{nd}$ ed</u>. Translated by Joseph Szücs. Edited by Mark Levi. New York: Springer-Verlag, 1988.

For a discussion of deterministic versions of quantum mechanics, see:

Bohm, David. <u>Causality and Chance in Modern Physics</u>. London: Routledge and Kegan Paul, 1957.

Cushing, James T. <u>Quantum Mechanics: Historical Contingency and the Copenhagen Hegemony</u>. Chicago: University of Chicago Press, 1994.

Possible consequences of determinism for free will in terms of the consequence argument may be found in:

Kane, Robert, ed. <u>The Oxford Handbook of Free Will</u>. New York: Oxford University Press, 2002.

van Inwagen, Peter. <u>An Essay on Free Will</u>. Oxford: Clarendon Press, 1983.

For a discussion of difficulties in applying determinism as unique evolution to psychology, see:

Bishop, Robert C. "Deterministic and Indeterministic Descriptions." In <u>Between Chance and Choice: Interdisciplinary Perspectives on Determinism</u>. Edited by Harald Atmanspacher and Robert C. Bishop. Thorverton: Imprint Academic, 2002.

Elements of the faithful model assumption have received some scrutiny in recent physics literature. In particular, there is evidence that perfect models are not guaranteed to describe system behavior in nonlinear contexts:

Judd, Kevin, and Smith, Leonard. "Indistinguishable States I: Perfect Model Scenario."

   Physica D 151(2001):125-141.
Judd, Kevin, and Smith, Leonard. "Indistinguishable States II: Imperfect Model Scenarios."
   Physica D 196(2004):224-242.
Smith, Leonard A. "Disentangling Uncertainty and Error: On the Predictability of Nonlinear
   Systems." In Nonlinear Dynamics and Statistics. Edited by A. I. Mees. Boston:
   Birkäuser, 2001.